# IS THE UNIVERSE REALLY EXPANDING?

Walter Petry
Mathematisches Institut der Universitaet Duesseldorf, D-40225 Duesseldorf
E-mail: wpetry@meduse.de
petryw@uni-duesseldorf.de

Abstract: The redshift of galaxies is in general explained by the expansion of space. The flat space time theory of gravitation suggests an additional interpretation. In this theory gravitation is explained analogously to Maxwell's theory on a flat space-time metric and gravitation is described by a field (potentials) with which the proper time (atomic time) is defined. In addition to the proper time in the universe, the oberserver's time is stated. The oberserver's time interval is absolute whereas the interval of the proper time is time dependent. In particular, atomic clocks at distant objects are going slower than clocks at present. This explains the redshift of distant objects without assuming an expanding universe.

### 1. Introduction

Hubble discovered since the late of 1920 for nearby galaxies a linear relation between the distance to the galaxy and the redshift of its spectrum. The constant of this relation is the well known Hubble constant. This redshift is generally explained as a Doppler frequency shift. This implies on the basis of a homogeneous, isotropic universe by Einstein's theory an expansion of space. Hubble had doubts about the reality of the expansion whereas Sandage and collaborators believe that the expansion of the universe is a reality. For a discussion of these two different meanings compare the paper of Soares [1] where further references on this subject can be found. The interpretation of the redshift as an expansion of the universe is generally accepted but there exist also some problems with the interpretation of an expanding space, e.g. the superluminal recession velocities of distant galaxies ( see e.g. [2-4]).

In this paper we start from a covariant theory of gravitation in flat space-time given by Petry [5]. This theory has been studied in several papers and it gives the same results as Einstein's general theory of relativity to the experimentally needed accuracy for the following effects: gravitational redshift, light deflection, perihelion precession, radar time-delay, post-Newtonian approximation, gravitational radiation and the precession of the spin axis of a gyroscope in the orbit of a rotating body. But Birkhoff`'s theorem is not valid and the theory gives under natural assumptions nonsingular, homogeneous, isotropic cosmological models in contrast to general relativity. A summary of these results and further references can be found in the paper of Petry [6]. Nonsingular, homogeneous, isotropic cosmological models are studied in several papers (see e.g. Petry [6-11] ). The space of flat space-time theory of gravitation is flat for all cosmological models as recent observations suggest. The used flat space-time metric of the considered cosmological models is the pseudo-Euclidean geometry where all covariant derivatives are the classical derivatives. In addition to this time and the proper time (atomic time) the time of the observer is defined ( see Petry [12] ). The proper time interval and the observer's time interval are different at distant objects and they only agree at present. This result has been used to give an explanation of the anomalous Doppler frequency shift of the Pioneers [12] which had been stated by many authors ( see e.g. Anderson et al. [13] ). It is shown that the observer's time interval is absolute implying that the invariant proper time interval is changeable in the universe. Therefore, clocks at earlier times measured by proper time are going slower than at present. The energy of matter in the universe decreases whereas the total energy is conserved implying this result. This is in agreement with the known result that clocks are going slower in the gravitational field of a static spherically symmetric body. The difference between the absolute time of the observer and the variable proper time of distant objects in the universe gives an explanation of the observed redshift of distant galaxies. Hence, the redshift is explained without the assumption of the expansion of space.

### 2. Summary of some Results

The theory of gravitation in flat space-time [5] has been studied in several papers. A summary of this theory with several applications can be found in [6] where references to the detailed studies are stated. Subsequently, we summarize some results of the papers [5-11] which are used in the following.

The flat space-time theory of gravitation starts with a flat space-time background metric

$$(ds)^2 = -\eta_{ij} dx^i dx^j. \tag{2.1}$$

The gravitational field is described by a symmetric tensor $g_{ij}$ satisfying covariant (with regard to the flat space-time metric (2.1)) differential equations of order two where the source of the gravitational field is the total energy-momentum tensor inclusive that of the gravitational field which is a tensor. The proper time $\tau$ (atomic time) is defined by

$$c^2 (d\tau)^2 = -g_{ij} dx^i dx^j. \tag{2.2}$$

The equations of motion of a test particle with the four-velocity $\left(\dfrac{dx^i}{d\tau}\right)$ in the gravitational field are give by

$$\frac{d}{d\tau}\left(g_{ij}\frac{dx^j}{d\tau}\right)=\frac{1}{2}\frac{\partial g_{jk}}{\partial x^i}\frac{dx^j}{d\tau}\frac{dx^k}{d\tau} \quad (i=1-4). \tag{2.3}$$

The total energy-momentum is conserved. The derivation of these results can be found in the paper of Petry [5]. These results are not needed subsequently and therefore, they are omitted here.

The application of the theory of gravitation in flat space-time to homogeneous, isotropic cosmological models is studied in several papers (see e.g. Petry [6-10] ). It starts with the pseudo-Euclidean geometry

$$(\eta_{ij}) = diag(1,1,1,-1) \tag{2.4}$$

as background metric where $x^1, x^2, x^3$ are the Cartesian coordinates and $x^4 = ct$. The four-velocity of the universe is

$$u^i = 0 \ (i=1-3), \quad u^4 = c\frac{dt}{d\tau}. \tag{2.5}$$

Then, the potentials $(g_{ij})$ are given by

$$(g_{ij}) = diag(a^2(t), a^2(t), a^2(t), -1/h(t)) \tag{2.6}$$

where $a(t)$ and $h(t)$ satisfy two coupled nonlinear differential equations of order two. Here, the source of the gravitational field are the densities of matter (dust) $\rho_m(t)$, of radiation $\rho_r(t)$ and of the cosmological constant $\rho_\Lambda(t)$ with

$$\rho_m = \frac{\rho_{m0}}{\sqrt{h(t)}}, \quad \rho_r = \frac{\rho_{r0}}{a(t)\sqrt{h(t)}}, \quad \rho_\Lambda = \frac{\Lambda c^2}{8\pi k}\frac{a^3(t)}{\sqrt{h(t)}}. \tag{2.7}$$

The parameters $\rho_{m0}$ and $\rho_{r0}$ are the densities at present, $\Lambda$ is the cosmological constant and $k$ the gravitational constant. In some models an additional field is considered with a stiff equation of state.

The initial conditions of the differential equations are given at present time $t=0$, i.e.

$$a(0) = h(0) = 1, \quad \dot{a}(0) = H_0, \quad \dot{h}(0) = \dot{h}_0. \tag{2.8}$$

Here, the dot denotes the time derivative, $H_0$ is the Hubble constant and $\dot{h}_0$ is a further constant of integration which is zero for Einstein's theory because $h(t) \equiv 1$ being not possible in flat space-time theory of gravitation. It follows that under natural conditions there exist nonsingular, cosmological models, i.e. all the energies are finite in contrast to Einstein's general theory of relativity. The functions $a(t)$ and $h(t)$ are defined for all $t \in R$. The function $a(t)$ starts from a small positive value as $t \to -\infty$, decreases a little to a small positive value and then increases for all $t \in R$. The function $h(t)$ starts from infinity as $t \to -\infty$, decreases to a small positive value and then increases to infinity as $t \to \infty$. The minimal values of $a(t)$ and $h(t)$ are reached in the very early universe. The energy density of the gravitational field is

$$\rho_G(t) = \frac{1}{64\pi k}\left(-6\left(\frac{\dot{a}}{a}\right)^2 + 6\frac{\dot{a}}{a}\frac{\dot{h}}{h} + \frac{1}{2}\left(\frac{\dot{h}}{h}\right)^2\right). \tag{2.9}$$

The space of the universe is flat for all cosmological models which is also suggested by recent observations. All these results can be found in the above specified papers of Petry on cosmological models. It is worth mentioning that the total energy of the universe is constant and non zero, i.e.

$$(\rho_m(t) + \rho_r(t) + \rho_\Lambda(t) + \rho_G(t))c^2 = \lambda c^2 \tag{2.10}$$

with a constant $\lambda \neq 0$, whereas Einstein's theory implies for a flat space that the total energy is zero ( see e.g. Berman [14] ).

### 3. Different Times and Frequencies

In addition to the time $t$ of the universe we have the proper time $\tilde{\tau}$ at time $t$ defined by an observer at rest by (see (2.2) with (2.6) )

$$d\tilde{\tau} = dt/\sqrt{h(t)} \tag{3.1}$$

with

$$\tilde{\tau}(t) = \int_{-\infty}^{t} \frac{dt}{\sqrt{h(t)}} \tag{3.2}$$

implying the potentials

$$(g_{ij}(\tilde{\tau})) = diag(a^2(t), a^2(t), a^2(t), -1) \tag{3.3}$$

Here, $t$ must be replaced by $\tilde{\tau}$ by the use of the inverse of relation (3.2).

The observer's time $t'$ is defined by (see Petry [11])

$$dt' = dt/(a(t)\sqrt{h(t)}) \tag{3.4}$$

with

$$t'(t) = \int_{-\infty}^{t} dt/(a(t)\sqrt{h(t)}) \tag{3.5}$$

implying the potentials

$$(g_{ij}(t')) = (a^2(t), a^2(t), a^2(t), -a^2(t)) \tag{3.6}$$

Here, $t$ must be replaced by $t'$ by the use of (3.5).

The light velocity $v_L$ with regard to the three different times $t, \tilde{\tau}$ and $t'$ in the universe at time $t$ is given by

$$v_L(t) = c/(a(t)\sqrt{h(t)}), \quad \tilde{v}_L(t) = c/a(t), \quad v'_L(t) = c. \tag{3.7}$$

Hence, the light velocity in the universe with observer's time is the constant light velocity $c$ independent of the time.

It follows by the initial conditions (2.8) and (3.1) resp. (3.4) that at present time $t = 0$

$$dt = d\tilde{\tau} = dt'. \tag{3.8}$$

Therefore, the frequency of a photon emitted at present by an atom at rest is by (3.8) independent of the used times $t, \tilde{\tau}$ or $t'$.

Let us calculate the frequency of a photon emitted by an atom in a distant object at rest and moving in the universe. The emitted energy of this photon is given by

$$E \sim -g_{44} \frac{dt}{d\tau}. \tag{3.9}$$

Hence, it follows for an atom at rest emitting at time $t_e$ a photon by the use of (2.2) and (2.6), (3.3) and (3.6) for $t, \tilde{\tau}$ and $t'$

$$E(t_e) = \frac{1}{h(t_e)} \sqrt{h(t_e)} E_0 = \frac{1}{\sqrt{h(t_e)}} E_0 \tag{3.10a}$$

$$\tilde{E}(t_e) = 1 \cdot E_0 = E_0 \tag{3.10b}$$

$$E'(t_e) = a^2(t_e) \frac{1}{a(t_e)} E_0 = a(t_e) E_0 \tag{3.10c}$$

where $E_0$ is the energy of the photon emitted by the same atom at rest and at present. The energy of a photon moving in the universe follows by the use of the equations of motion (2.3) with $i = 4$. We get by the use of the time $t'$ and (3.6)

$$\frac{d}{dt'}\left(-a^2(t)\frac{dt'}{d\tau}\right) = \frac{1}{c^2} a(t)\dot{a}(t)(|v'_L(t)|^2 - c^2)\frac{dt}{dt'}\frac{dt'}{d\tau} = 0. \tag{3.11}$$

Here, $|\ |$ denotes the Euclidean norm and relation (3.7) is used.

Therefore, it follows for all times

$$E'(t) \sim a^2(t)\frac{dt'}{d\tau} = const.$$

implying by the use of (3.10c) for all $t$

$$E'(t) = E'(t_e) = a(t_e)E_0. \tag{3.12}$$

This relation gives by Planck's law for the frequency
$$v'(t) = a(t_e)v_0 \qquad (3.13)$$
where $v'(t)$ is the frequency of the moving photon at time $t$ and $v_0$ is the frequency of the photon emitted by the same atom at rest and at present. The wavelength of the photon follows by the use of
$$g^{ij} p_i p_j = 0$$
implying
$$\frac{1}{a^2(t)}\left(|p(t)|^2 - p_4^2(t)\right) = 0$$
where $p(t)$ is the momentum of the photon. We get by Planck's law and (3.13)
$$\lambda'(t) = \frac{c}{a(t_e)} \frac{1}{v_0} \qquad (3.14)$$
for all times. Hence, it follows
$$\lambda'(t)v'(t) = c \qquad (3.15)$$
which is by (3.7) the light velocity at all times measured by the use of the observer's time $t'$.

The equations (3.13) and (3.14) show that by the use of observer's time the frequency and the wavelength of a photon emitted at a distant object are constant on the way moving to the observer but the emitted frequency and wavelength are different from those emitted by the same atom at present. It follows by (2.2) and (3.6) that the proper time $\tilde{\tau}$ of an object at rest and at time $t$ is given by
$$d\tilde{\tau} = a(t)dt'. \qquad (3.16)$$
Furthermore, we get by the use of (2.2) and (3.6) that two light signals emitted from a distant object at rest with time interval $dt'$ are received by the observer with the same time interval $dt'$, i.e. $dt'$ is not changed and it is equal to the present atomic time interval $d\tilde{\tau}$. This result does not hold if we use the time interval $dt$ or $d\tilde{\tau}$. The interval $d\tilde{\tau}$ is an invariant quantity for any object at rest and for all the times $t'$ implying by (3.16) that $d\tilde{\tau}$ depends on the time, i.e. in relation (3.16) it holds $d\tilde{\tau} = d\tilde{\tau}(t)$. Hence, $dt'$ is the absolute time interval whereas $d\tilde{\tau}(t)$ is not absolute but depends on the time in the universe. Therefore, atomic clocks in the universe are going slower in earlier times than at present by virtue of (3.16). This result can be explained by the remark that the total energy of the universe is conserved but the energy of matter and of gravitation are tranformed into one another in the course of time by virtue of (2.7), (2.9) and (2.10). In particular, in earlier times the density of matter was higher than at present. This result on clocks in the universe is in agreement with clocks in a stationary gravitational field produced by a mass where clocks are going slower in the neighborhood of the mass. The equations (3.13) and (3.16) are equivalent since the energy of a photon in the universe is not changed. Relation (3.13) gives the well known redshift formula of distant galaxies measured by the observer at present time $t = 0$:
$$z = \frac{v_0}{v'(0)} - 1 = \frac{1}{a(t_e)} - 1. \qquad (3.17)$$
Hence, the redshift of an object at rest in the universe follows by the result that the time interval of the proper time $d\tilde{\tau}(t)$ is time dependent whereas the observer's time interval $dt'$ is the same for all the times and it holds (3.16). Therefore, the redshift is explained without the assumption of an expanding space.

It is worth mentioning that relation (2.2) with (3.6) defines the proper time and relation (2.1) with (3.4) the line element of the universe whereas in Einstein's theory the two relations are identical and given by (2.2). This result and the density of matter (2.7) also indicate that the space of the universe is not expanding.

Formula (3.16) implies that the duration of a process in a distant object at rest and measured with atomic time is seen by the observer with longer time interval and therefore, it is longer than the duration of the same process taking place at present. This is in agreement with the observed duration of the brightness of supernovae.

Along the lines of the above considerations the anomalous Doppler frequency shift of the Pioneers has been explained by Petry [12].

It is worth mentioning that the frequency and the wavelength of the moving photon emitted from an atom in a distant object at rest also can be received by the use of the time $t$ or $\tilde{\tau}$ along the lines of these considerations. The time $t$ was considered in paper [8]. It holds
$$v(t) = \frac{a(t_e)}{a(t)\sqrt{h(t)}} v_0, \quad \lambda(t) = \frac{c}{a(t_e)} \frac{1}{v_0} \qquad (3.18)$$
implying the light velocity (see (3.7)) at time $t$

$$\nu(t)\lambda(t)=\frac{c}{a(t)\sqrt{h(t)}} \quad . \tag{3.19}$$

We get by the use of $\tilde{\tau}$

$$\tilde{\nu}(t)=\frac{a(t_e)}{a(t)}\nu_0 \; , \qquad \tilde{\lambda}(t)=\frac{c}{a(t_e)}\frac{1}{\nu_0} \tag{3.20}$$

implying again the light velocity (compare (3.7))

$$\tilde{\nu}(t)\tilde{\lambda}(t)=\frac{c}{a(t)} \quad . \tag{3.21}$$

The wavelengthes are by (3.14), (3.18) and (3.20) for $t, \tilde{\tau}$ and $t'$ the same and constant during the motion of the photon whereas the frequency measured with time $t$ or $\tilde{\tau}$ is time dependent during the motion of the photon but at present time it is by virtue of the initial conditions (2.8) identical with the frequency (3.13) measured with the observer's time $t'$.

It is interesting that Zakir [15] also concludes in considering black holes that the invariance of proper time does not mean its absoluteness as generally assumed.